\documentclass[usenatbib]{mnras}
\usepackage{graphicx}
\usepackage{natbib}
\usepackage{latexsym}
\usepackage{amsmath}
\usepackage{setspace}

\newcommand{\etal}{et~al.} 
\newcommand{\ionhy}{H{\sc ii} }

\newcommand{\kms}{$\mbox{km~s}^{-1}$}
\newcommand{\msol}{\mbox{M\hbox{$_{\odot}$}}}
\newcommand{\lsol}{\mbox{L\hbox{$_{\odot}$}}}
\newcommand{\ammonia}{$\mbox{NH}_{3}$}
\newcommand{\cyanoacet}{$\mbox{HC}_3\mbox{N}$}

\newcommand{\specsfig}[1]        % Double FIGures (put two figures  
{
   \begin{center}
     \begin{minipage}[t]{0.45\textwidth}
         \psfig{file=#1.eps,height=0.9\textwidth,angle=270}
     \end{minipage}
     \end{center}
 }

\newcommand{\specdfig}[2]        % Double FIGures (put two figures  
{
   \begin{center}
     \begin{minipage}[t]{0.45\textwidth}
         \psfig{file=#1.eps,height=0.9\textwidth,angle=270}
     \end{minipage}
     \hfill
     \begin{minipage}[t]{0.45\textwidth}
         \psfig{file=#2.eps,height=0.9\textwidth,angle=270}
     \end{minipage}
   \end{center}
}

\begin{document}

\title[Class~I methanol masers in NGC\,253]{Class~I methanol masers in NGC\,253: Alcohol at the end of the bar}
\author[Ellingsen \etal]{S.\ P. Ellingsen,$^{1}$\thanks{Email: Simon.Ellingsen@utas.edu.au} X. Chen,$^{2,3}$, S.\ L. Breen,$^{4}$, H-H. Qiao,$^{3,5}$ \\
  \\
  $^1$ School of Physical Sciences, University of Tasmania, Private Bag 37, Hobart, Tasmania 7001, Australia\\
  $^2$ Center for Astrophysics, GuangZhou University, Guangzhou 510006, China\\
  $^3$ Shanghai Astronomical Observatory, Chinese Academy of Sciences, Shanghai 200030, China\\
  $^4$ Sydney Institute for Astronomy (SIfA), School of Physics, University of Sydney, Sydney, NSW 2006, Australia\\
  $^5$ National Time Service Center, Chinese Academy of Sciences, Xi'An, Shanxi, 710600, China}

 \maketitle

\begin{abstract}
We have used the Australia Telescope Compact Array (ATCA) to observe the 36.2-GHz class~I methanol maser emission towards NGC\,253 and find that it is located at the interface between the nuclear ring and both ends of the galactic bar.  This is thought to be the location of the inner Linblad resonance and we suggest that the maser emission in this region is likely due to large-scale cloud-cloud collisions.  We have detected the first extragalactic 44.1-GHz class~I methanol maser and find that it is associated with the 36.2-GHz maser emission.  In contrast to the class~I methanol masers found in Galactic star formation regions, the 44.1-GHz emission in NGC\,253 is two orders of magnitude weaker than the 36.2-GHz masers.  Both the 36.2- and 44.1- GHz emission is orders of magnitude stronger than expected from typical high-mass star formation regions. This demonstrates that the luminous class~I methanol masers observed in NGC\,253 are significantly different from those associated with Galactic star formation.  
\end{abstract}


\section{Introduction}

NGC\,253 is a nearby \citep[3.4 Mpc;][]{Dalcanton+09}, barred-spiral galaxy in the Sculptor group, and is one of the closest galaxies with a nuclear starburst. It shows strong molecular emission from a wide range of species \citep[e.g.][]{Martin+06,Meier+15} and has been imaged at high angular resolution and sensitivity across the electromagnetic spectrum \citep[e.g.][]{Ulvestad+97,Sakamoto+11,Weiss+08,Iodice+14,Dale+09,Lehmer+13}.

The proximity of NGC\,253 makes it possible to study the nuclear starburst at high resolution and sensitivity and it provides an important testbed for understanding the relationship between star formation in normal spirals, nuclear starbursts and merger-driven starburst systems \citep{Leroy+15}.  The star formation rate within the nuclear starburst region is estimated to be approximately 2~\msol year$^{-1}$, with both radio \citep{Ott+05} and infrared \citep{Radovich+01} observations giving similar results.  NGC\,253 is a barred spiral and it is widely thought that the bar-potential plays a critical role in driving the nuclear starburst \citep[e.g.][]{Garcia-Burillo+00,Iodice+14}.  The relative intensities of the high-excitation states of CO in the central molecular zone (CMZ) require both ultraviolet and mechanical heating to explain the observations and while cosmic-ray heating may also contribute, it is not required \citep{Rosenberg+14}.

\citet{Iodice+14} show the presence of a nuclear ring at approximately the radius of the inner Linblad resonance, representing the inner end of the bar.  The star formation in the central few hundred parsecs of NGC\,253 (within the nuclear ring) accounts for about 50 percent of the total star formation within the galaxy \citep{Leroy+15}. The radio and molecular emission from NGC\,253 essentially all arises within the nuclear ring, with the radio emission being predominantly due to supernovae and \ionhy\/ regions \citep{Ulvestad+97,Lenc+06}. The compact non-thermal radio source TH2 has been suggested as a possible Sgr A$^*$-like central supermassive blackhole \citep{Ulvestad+97}, however, the absence of an x-ray counterpart \citep{Lehmer+13} and milliarcsecond-scale radio emission \citep{Brunthaler+09} casts doubt on this interpretation \citep{Muller-Sanchez+10}.

The galaxy has long been known to host OH emission \citep{Turner+85}, water masers \citep{Henkel+04,Hofner+06} and an \ammonia(3,3) maser is detected toward the centre \citep{Ott+05,Gorski+17}.  The first detection of extragalactic class~I methanol maser emission from the 36.2-GHz $4_{-1} \rightarrow 3_{0}E$ transition was reported towards NGC\,253 \citep{Ellingsen+14} and most recently a \cyanoacet\/ maser from the J = 4--3 transition at 36.4~GHz has been detected close to the class I methanol masers \citep{Ellingsen+17}.  The discovery of the 36.2-GHz methanol emission was made at modest spatial and spectral resolution and has recently been confirmed through independent Jansky Very Large Array (JVLA) observations \citep{Gorski+17}.  In this paper we report new observations of NGC\,253 at 36 and 44~GHz with the Australia Telescope Compact Array (ATCA) which we have used to investigate the nature of the methanol emission and its relationship with the dynamics and other properties of the galaxy.

\section{Observations} \label{sec:observations}

The observations were made using the Australia Telescope Compact Array (ATCA) on 2014 October 10, 11 and November 27, 28 (project code C2879).  The array was in the 1.5A configuration (baseline lengths between 153 and 1469 m) for the October session and in the EW367 (baseline lengths between 46 and 367 m) for the November session.  The current observations are sensitive to a maximum angular scale of approximately 15 arcseconds and 53 arcseconds for the 1.5A and EW367 array observations respectively.  The Compact Array Broadband Backend \citep{Wilson+11} was configured with 2 $\times$ 2.048 GHz bands, both centred on the same frequency.  One of the 2.048 GHz bands had 2048 spectral channels each of 1 MHz, while the other had between two and five ``zoom-bands'' with spectral channels of width 31.25~kHz and bandwidths which ranged from 128 to 224~MHz (corresponding to 4096 -- 7168 channels per zoom-band). For each array configuration we made observation centred at a frequency of 36.9~GHz on one day and at 43.3~GHz on the other day.  The 36.9-GHz observations cover the rest frequency of the $4_{-1} \rightarrow 3_{0}E$ (36.2-GHz class~I) and $7_{-2} \rightarrow 8_{-1}E$ (37.7-GHz class~II) methanol transitions on one day and of the $7_{0} \rightarrow 6_{1}A^{+}$ (44.1-GHz class~I) methanol transition on the other day.  We adopted rest frequencies of 36.169265, 37.703700 and 44.069410~GHz respectively for the three methanol maser transitions \citep{Muller+04}. The $v$ = 0, 1, 2 \& 3 transitions of SiO (1--0) were observed simultaneously with the 44.1-GHz methanol transition.  Details of the observations, including the resulting synthesised beam and image noise levels are summarised in Table~\ref{tab:obs}.  The primary beam of the ATCA antennas is approximately 1.5 arcminutes at 36.2~GHz and 1.2 arcminutes at 44.1~GHz.  All the detected methanol emission lies well within the primary beam of the observations and we have not undertaken any scaling of the reported flux densities to correct for the primary beam.

The current observations used the same primary calibrator (Uranus) and phase calibrator (0116-219) as the March 2014 observations made in the H168 array reported by \citet{Ellingsen+14}.  From hereon we refer to the different ATCA observations as high-, intermediate- and low-resolution for the data taken in the 1.5A, EW367 and H168 arrays respectively.  The low-resolution observations were less sensitive (30 minutes onsource) and lower spectral resolution (8.2~\kms) than the intermediate- and high-resolution observations reported here.  Comparing the peak flux density of the 36.2-GHz continuum emission from NGC\,253 and that determined for the phase calibrator between the low- and intermediate-resolution observations we find a ratio very close to 0.5:1 for both sources.  The three observing sessions are separated by a period of 8 months, so some variability in the flux density of the phase calibrator is possible. Undertaking the same comparison between the high- and intermediate-resolution observations of the phase calibrator (0116-219), we measure a ratio of 0.88:1 for these data taken 6 weeks apart.  We also examined the amplitude of the 0116-219 as a function of baseline length for each of the array configurations and it shows constant amplitude for each of them. 

The data reduction followed the same procedure as for \citet{Ellingsen+17}.  Image cubes of the maser and thermal lines were created from the continuum subtracted, self-calibrated data.  The intrinsic velocity resolution of the zoom-band observations of the maser transitions (31.25-kHz spectral resolution) ranges from 0.25--0.32~\kms, while for the 1-MHz spectral resolution data the velocity resolution ranges from 6.8--8.2~\kms. We found image cubes with a velocity resolution of 3~\kms\/ provided the best compromise between resolution and signal to noise in individual velocity channels for the maser emission.   For the stronger thermal transitions the intrinsic velocity resolutions had adequate signal to noise, while for the weaker transitions we averaged to a final resolution of 20~\kms.  We used the barycentric reference frame for the velocities in all image cubes and {\sc miriad} corrects for the changing line-of-sight velocity of the observatory with respect to the source prior to imaging.  All data have been self-calibrated using the continuum data from the central region of the galaxy, which effectively aligns the observations from different epochs and frequencies to the degree to which the peak of the continuum emission is the same between the 36- and 44-GHz observations and the relative astrometric accuracy is expected to be a fraction of an arcsecond. 

\begin{table*}
\caption{Summary of observations}
  \begin{tabular}{lccccccc} \hline
        &  &  &  & \multicolumn{2}{c}{\bf Velocity} & \\
      \multicolumn{1}{c}{\bf Date} & \multicolumn{1}{c}{\bf Array}  & \multicolumn{1}{c}{\bf Centre} & \multicolumn{1}{c}{\bf Time} & \multicolumn{1}{c}{\bf Width} & \multicolumn{1}{c}{\bf Resolution}  & \multicolumn{1}{c}{\bf Synthesised} & \multicolumn{1}{c}{\bf Image RMS}  \\
                                                        & \multicolumn{1}{c}{\bf Configuration} & \multicolumn{1}{c}{\bf Frequency (GHz)} &  \multicolumn{1}{c}{\bf Onsource (min)} & \multicolumn{1}{c}{\bf (\kms)} & \multicolumn{1}{c}{\bf  (\kms)} & \multicolumn{1}{c}{\bf Beam} & \multicolumn{1}{c}{\bf (mJy beam$^{-1}$)} \\ \hline \hline
 10 Oct 2014 & 1.5A & 36.9~GHz & 296 & -250 -- 750 & 3.0 &  $0.8 \times 3.5$ & 1.1 \\
 11 Oct 2014 & 1.5A & 43.3~GHz & 256 & -125 -- 700 & 3.0 & $0.5 \times 2.6$ & 1.1 \\
 27 Nov 2014 & EW367 & 36.9~GHz & 232 & -265 -- 735 & 3.0 & $3.1\times 16.6$ & 1.1 \\
 28 Nov 2014 & EW367 & 43.3~GHz & 315 & -140 -- 680 & 3.0 & $2.6\times 12.2$ & 1.6 \\ \hline \hline
\end{tabular} \label{tab:obs}
\end{table*}

\section{Results} \label{sec:results}

\subsection{36.2-GHz methanol}

\begin{figure}
   \includegraphics[scale=0.35,angle=270]{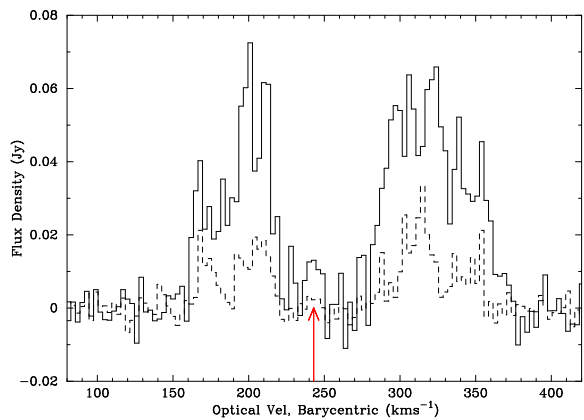}
   \caption{The integrated 36.2-GHz methanol emission from NGC\,253 from the ATCA intermediate-resolution observations (solid line) and high-resolution observations (dashed line).  The red arrow marks the systemic velocity of the galaxy (243 $\pm$ 2) \kms\/}
  \label{fig:integrated}
\end{figure}

We detected the 36.2-GHz ($4_{-1} \rightarrow 3_{0}E$) methanol transition in both ATCA array configurations and Figure~\ref{fig:integrated} shows the integrated emission.   The maximum baseline length for the high-resolution observations is nearly an order of magnitude longer than the previous low-resolution observations of this emission \citep{Ellingsen+14}.  The integrated emission from the 36.2-GHz transition is 5.8 Jy~\kms\/ for the intermediate-resolution data and 1.4 Jy~\kms\/ for the high-resolution data.  So three quarters of the 36.2-GHz methanol emission is resolved in the high-resolution observations, suggesting that it is on scales larger than about 15 arcseconds (corresponding to a linear scale of 250 parsecs).  The integrated flux density reported by \citet{Ellingsen+14} for the low-resolution observations was 0.6 Jy~\kms\/, whereas we would expect it to be equal to, or larger than that measured for the intermediate-resolution array data.  If we assume that this difference is due to either decorrelation or poor flux density calibration in the low-resolution observations and that they should be scaled by a factor of 2 (see section~\ref{sec:observations}), that produces an integrated spectrum from the low-resolution data which shows better agreement with the intermediate-resolution data in the peak amplitude across much of the spectrum.  However, it still results in an integrated flux density more than a factor of 4 lower than that observed in the intermediate-resolution observations, predominantly because the low-resolution data does not exhibit the emission around 300 and around 330~\kms\/ seen in the intermediate resolution spectrum (compare figure~2 of \citet{Ellingsen+14} with Figure~\ref{fig:integrated}).   The combination of differing array configurations and spectral resolution, in addition to the uncertainties related to primary flux density calibration, phase calibration and possible temporal variation in the phase calibrator mean that it is not possible to definitively determine if the 36.2-GHz methanol emission has shown temporal variability, however, that is certainly a possibility, especially because the significant changes in the spectral profile are difficult to explain any other way.

\begin{figure*}
   \includegraphics[scale=0.35]{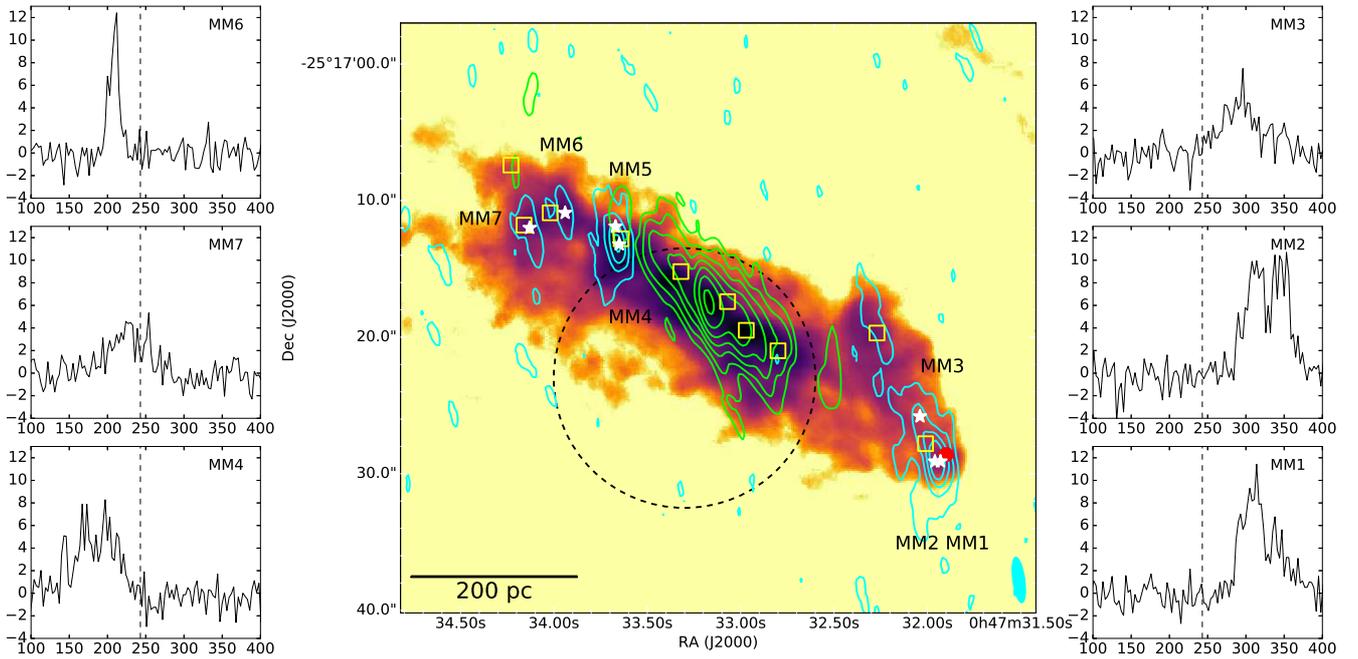}
   \caption{{\em Centre panel:} ATCA image of the integrated 36.2-GHz methanol emission (cyan contours at 0.24, 0.34, 0.44 and 0.54 Jy \kms\/ beam$^{-1}$) and 7-mm continuum emission (green contours at 1.25, 2.5, 5, 10, 20, 40 and 80\% of 100 mJy~beam$^{-1}$)  towards NGC\,253.  The synthesised beam for the ATCA methanol observations is $0.9 \times 3.4$ arcseconds at a position angle of 8 degrees (shown in the bottom-right corner of image) and combines the high-, intermediate and low-resolution data. The background image is the integrated CO J = 2--1 emission from \citet{Sakamoto+11} shown on a logarithmic scale. The yellow squares mark the location of molecular clouds identified by \citep{Leroy+15} and the red circle the location of the 36.4-GHz \cyanoacet\/ maser  \citep{Ellingsen+17}.  The white stars mark the locations of the seven methanol maser sites identified in the current observations, which we have labelled MM1 through MM7 by increasing right ascension.  The dashed black circle marks the pointing centre and size of the full-width at half-maximum (FWHM) primary beam of the IRAM 30-m observations by \citep{Martin+06}.  The FWHM primary beam of the current ATCA observations is larger than the size of the image.  {\em Side panels:} The spectra extracted from the ATCA 36.2-GHz methanol image cube (3~\kms\/ spectral channels) at the location of the labelled white stars in the centre panel.  The vertical dashed line marks the systemic velocity of NGC\,253 \citep[243~\kms][]{Koribalski+04}.  The x-axis for the spectra is the Barycentric velocity in \kms\/ and the y-axis is the flux in mJy beam$^{-1}$.}
  \label{fig:ngc253}
\end{figure*}

\begin{figure}
  \begin{center}
   \includegraphics[scale=0.5]{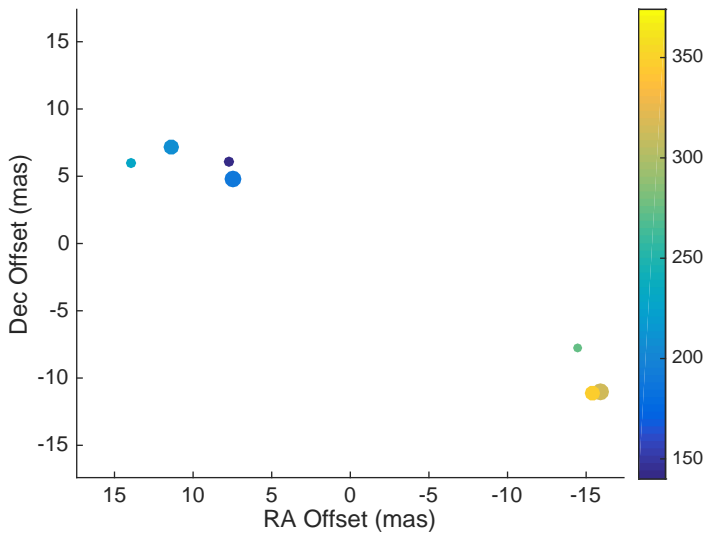}
   \caption{The distribution of 36.2-GHz methanol emission in NGC\,253 with respect to the pointing centre of the observations ($\alpha = 00^{\mbox{h}}47^{\mbox{m}}33.10^{\mbox{s}}$ ; $\delta = -25^\circ 17\arcmin18.0\arcsec$ (J2000)).  The size of the circle is proportional to the peak flux density and the colour indicates the Barycentric velocity (\kms).  }
  \label{fig:spot}
  \end{center}
\end{figure}

Figure~\ref{fig:ngc253} shows the 36.2-GHz emission formed by combining the data from all three available array configurations (the high- and intermediate resolution data reported here and the low-resolution data from March 2014).  The combined array image has a synthesised beam of $0.9 \times 3.4$ arcseconds and with a 3~\kms\/ velocity resolution the 36.2-GHz image cube has an RMS noise level of 1.1 mJy beam$^{-1}$.  The basic spatial and spectral morphology of the emission matches the earlier H168 array observations \citep[see figure~1 of][]{Ellingsen+14}.  There are two primary sites of methanol emission offset approximately 8 arcseconds to the north-east and 20 arcseconds to the south-west of the peak of the continuum emission.  Figure~\ref{fig:ngc253} shows that the new observations partially resolve the emission in both regions and the greater sensitivity reveals several additional sites of weaker emission.   To better characterise the maser regions we used the {\sc miriad} task {\tt imfit} to fit 2-dimensional Gaussian profiles for each plane of the image cube.  We then grouped maser emission with spatial separations less than 2 arcseconds in two or more consecutive velocity channels into single components, represented by the flux-density-weighted average position.  The locations and velocities of the components are shown in Figure~\ref{fig:spot} and their properties are summarised in Table~\ref{tab:component}.  We have labelled the seven components identified through this process MM1 through MM7 ordered by increasing right ascension (Fig.~\ref{fig:ngc253}).  The spectra extracted from the image cube are shown in the side panels of Figure~\ref{fig:ngc253} for six of the seven components (the emission from MM5 is not shown, but shows emission over a similar velocity range to MM4, but peaking at a lower velocity).  

\begin{table*}
\caption{Summary of the seven 36.2-GHz methanol maser components in NGC\,253 and their relationship to other molecular emission.  The values in parentheses in the last three columns are the offsets of the maser positions from those objects in arcseconds}
  \begin{tabular}{lccccccc} \hline
      \multicolumn{1}{c}{\bf Label} & \multicolumn{1}{c}{\bf Right Ascension}  & \multicolumn{1}{c}{\bf Declination} & \multicolumn{2}{c}{\bf Barycentric Velocity} & \multicolumn{1}{c}{\bf Gorski et al.} & \multicolumn{1}{c}{\bf Leroy et al.}& \multicolumn{1}{c}{\bf \cyanoacet} \\
                                                     & \multicolumn{1}{c}{\bf (J2000)}               & \multicolumn{1}{c}{\bf (J2000)}      &  \multicolumn{1}{c}{\bf Peak (\kms)}  & \multicolumn{1}{c}{\bf Range (\kms)} & \multicolumn{1}{c}{\bf label} & \multicolumn{1}{c}{\bf cloud} & \multicolumn{1}{c}{\bf source} \\ \hline \hline
MM1 & 00:47:31.93 & -25:17:29.1 & 314.0 & 287--347 & M5 (0.45) & 1 (1.7) & A (0.8) \\ % 301.3 (84.3)
MM2 & 00:47:31.96 & -25:17:29.1 & 345.5 & 290--362 &                 & 1 (1.4) & A (0.5) \\
MM3 & 00:47:32.04 & -25:17:25.8 & 275.0 & 272--305 & M4 (0.55) & 1 (2.0) &              \\ % 301.5 (35.7), 338.84 (54.5) ; 11.46 arcsec from C
MM4 & 00:47:33.65 & -25:17:13.2 & 188.0 & 140--221 & M3 (0.69) & 7 (0.4) & \\ % 173.1 (85.8)
MM5 & 00:47:33.67 & -25:17:11.9 & 143.0 & 140--224 &                 & 7 (1.0) & E (0.8) \\
MM6 & 00:47:33.94 & -25:17:10.9 & 207.5 & 200--215 & M2 (0.55) & 8 (1.1) & F (1.3) \\ % 202.6 (32.2)
MM7 & 00:47:34.13 & -25:17:12.0 & 228.5 & 212--254 & M1 (0.51) & 9 (0.5) & G (0.7) \\ \hline \hline % 209.2 (49.2) 
\end{tabular} \label{tab:component}
\end{table*}

\begin{figure*}
   \includegraphics[scale=0.35]{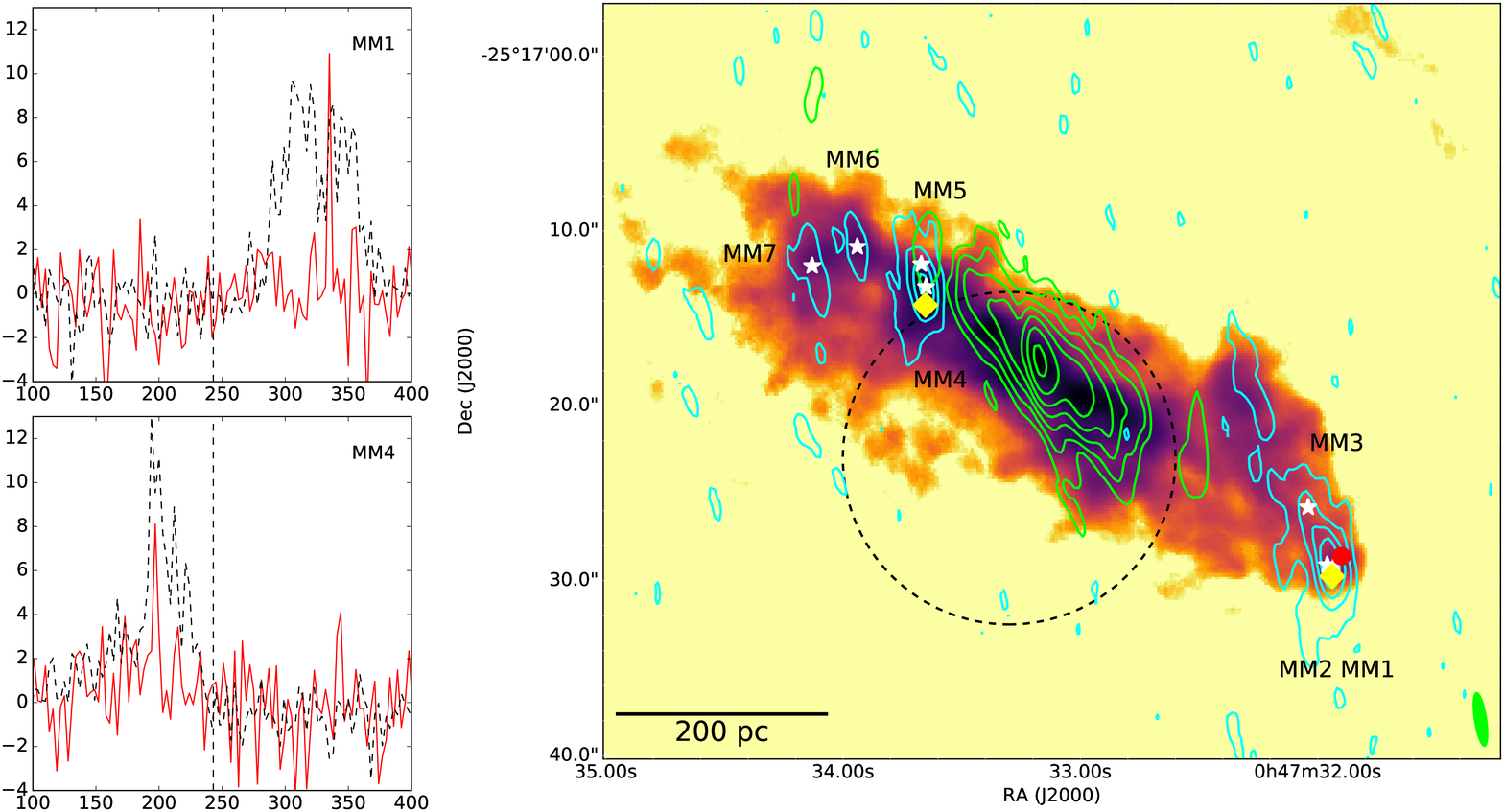}
   \caption{{\em Right panel:} The cyan diamond squares mark the location of 44-GHz methanol maser emission.  The synthesised beam of the ATCA 44.1-GHz image formed by combining data from the intermediate and high-resolution observations is 0.7 $\times$ 3.12 arcseconds at a position angle of 8.8 degrees and is shown in the bottom-right corner of the image.  The other information is as for the centre panel of Figure~\protect{\ref{fig:ngc253}}.  {\em Left panels:} The solid red lines are the spectra extracted from the ATCA 44.1-GHz methanol image cube (3~\kms\/ spectral channels) at the location of the labelled yellow diamonds in the centre panel.   The dashed black lines are the spectra extracted from the ATCA 36.2-GHz methanol image cube at the same location. The vertical dashed line marks the systemic velocity of NGC\,253.  The x-axis for the spectra is the Barycentric velocity in \kms\/ and the y-axis is the flux in mJy beam$^{-1}$.}
  \label{fig:newmeth}
\end{figure*} 

\subsection{44.1-GHz methanol}

These observations represent the first search for 44.1-GHz class~I methanol masers in NGC\,253 and emission is detected towards the MM1 and MM4 36.2-GHz methanol maser sites in both the intermediate- and high-resolution data.  The emission from the 44.1-GHz ($7_{0} \rightarrow 6_{1}A^{+}$) methanol transition shows a significantly narrower line width than the 36.2-GHz emission from the same location and the integrated intensity is more than an order of magnitude weaker (0.06 Jy~\kms).  Figure~\ref{fig:newmeth} shows the 44.1-GHz methanol emission towards NGC\,253.  The image was formed by combining data from the intermediate- and high-resolution observations.  The spectral line image cube for the 44.1-GHz methanol transition has an RMS of 1.75 mJy~beam$^{-1}$ and the emission towards locations MM1 and MM4 has a peak flux density 6.2 and 4.6 times this, respectively.  Statistically, the 44.1-GHz methanol detections are marginal.  However, the emission is observed at the same velocity and location in both array configurations and matches the velocity and location of stronger 36.2-GHz methanol emission.  We are therefore confident that this represents a robust detection of 44.1-GHz methanol emission.  The brightness temperature of the 44.1-GHz methanol emission towards MM1 is only 4.4~K, however, given the very narrow linewidth ($<$ 3~\kms) compared to that observed in dense gas tracers from this region there is little doubt that it is a maser.

\subsection{Other spectral lines}

\begin{figure*}
   \includegraphics[scale=0.30]{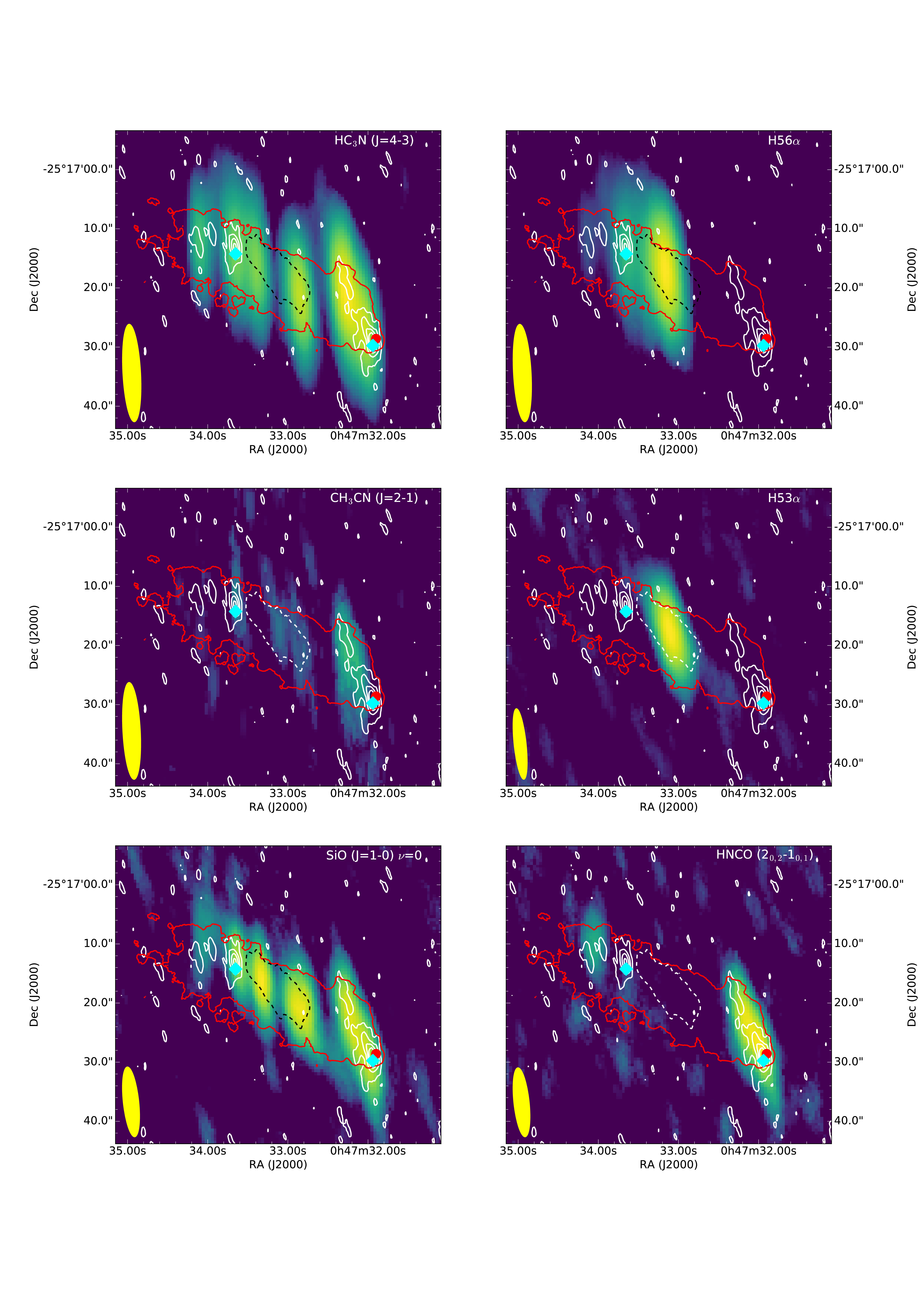}
   \caption{Molecular and atomic emission from transitions observed simultaneously at intermediate-resolution with the 36.2- and 44.1-GHz methanol transitions.  The colour image in each panel is the integrated emission of the molecular (or atomic) transition indicated in the top-right corner of the plot shown with on a logarithmic scale.  The synthesised beam for the molecular (or atomic) transition is shown in the bottom-left corner of the image as a yellow ellipse.  The red contour is the 5 percent contour for the integrated CO J = 2--1 emission \citep{Sakamoto+11}.  The white contours are the integrated 36.2-GHz methanol emission (at 0.24, 0.34, 0.44 and 0.54 Jy \kms\/ beam$^{-1}$).  The dashed contour (black in some panels, white in others) is the 2.5 percent contour for the 7-mm continuum emission.  The red circle marks the location of the 36.4-GHz \cyanoacet\/ maser \citep{Ellingsen+17} and the cyan diamonds the location of the 44.1-GHz methanol masers.}
  \label{fig:molecules}
\end{figure*}

A number of molecular and atomic transitions were covered by the 2-GHz IF band recorded for these observations, which have the advantage that they have the same angular resolution and sensitivity to the methanol observations, enabling a more direct comparison.  The details of the observations of these transitions are summarised in Table~\ref{tab:molecules}.  Figure~\ref{fig:molecules} shows the distribution of the observed molecular and atomic transitions from the intermediate-resolution observations.  The strongest and most widespread molecular transitions we observed were the \cyanoacet\/ J = 4--3 transition and the SiO J = 1--0 $\nu$ = 0, which span a similar spatial range to the CO J = 2--1 emission.  Both of these transitions show emission at all the 36.2-GHz methanol locations, as well as closer to the nucleus of the galaxy, where no methanol emission is detected.  Thermal SiO emission is a well-established extragalactic shock tracer \citep[e.g.][]{Garcia-Burillo+00}, while \cyanoacet\/ has a relatively high critical density and is observed in hot molecular core (HMC) sources such as Sgr\,B2 \citep[e.g.][]{Hunt+99}.  CH$_3$CN J = 2--1 is generally enhanced in HMC and in NGC\,253 it is observed to be strongest towards the south-western edge of the nucleus, but closer to the nucleus than the strongest 36.2-GHz methanol emission (Figure~\ref{fig:molecules}).  HNCO is a low-velocity shock tracer \citep{Meier+05}.  It is enhanced in regions where there is significant sublimation of dust ice mantle, and traces less energetic shocks than the those which produce strong SiO emission from sputtering of dust grain cores.  Figure~\ref{fig:molecules} shows that it is strongest towards the MM1-MM3 36.2 GHz methanol regions, with weaker emission in the MM6-MM7 region, but little or no emission in the MM4-MM5 complex.   

In addition to the four thermal molecular transitions, three recombination lines were detected, H56$\alpha$, H53$\alpha$ and H66$\beta$.  The first two of these are shown in Figure~\ref{fig:molecules}, showing a very different distribution from the molecular transitions.  H66$\beta$ shows a similar distribution, but is significantly weaker and so yields no useful additional information.  The peak in the recombination line intensity coincides with the peak of the continuum emission and it drops rapidly in intensity with distance from the nucleus.  Weaker H56$\alpha$ emission is detected towards the 36.2-GHz methanol sites east of the nucleus, but not in the MM1-MM3 complex.

\citet{Wang+14} have reported maser emission from the $5_{-1} \rightarrow 4_{0}E$ (84.5-GHz class~I) methanol and J = 2--1 $\nu$=3 (86~GHz) SiO transitions towards NGC\,1068, although to date there are no independent confirmations of either detection.  The two detected SiO J = 2--1 $\nu$=3 components have isotropic luminosities around 10$^7$ Jy~\kms\/kpc$^2$.  We have observed the SiO J = 1--0 $\nu$=0,1,2 and 3 transitions towards NGC\,253 but detect emission only from the vibrational ground state ($\nu$=0).  For the vibrationally excited states we constructed image cubes with 20~\kms\/ spectral resolution and these each have an RMS of better than 0.6 mJy beam$^{-1}$.  From this we can place an upper limit (using 5 times the RMS noise in the image cube) of 7 $\times$ 10$^5$ Jy~\kms\/kpc$^2$ on any  SiO J = 1--0 vibrationally excited-state maser emission in NGC\,253.  For comparison, this limit is about 200 times the luminosity of the SiO J = 1--0 $\nu$=1 maser emission from the Orion star formation region \citep{Goddi+09} and several thousand times the luminosity of typical late-type star SiO masers \citep{Deguchi+04}.

\begin{table*}
\caption{Molecular and atomic transitions detected in the 2-GHz IF band.}
  \begin{tabular}{lccccc} \hline
      \multicolumn{1}{c}{\bf Transition} & \multicolumn{1}{c}{\bf Rest}  & \multicolumn{1}{c}{\bf Array} & \multicolumn{1}{c}{\bf Synthesised } & \multicolumn{1}{c}{\bf Velocity} & \multicolumn{1}{c}{\bf Image RMS}  \\
                                                        & \multicolumn{1}{c}{\bf Frequency (GHz)} & \multicolumn{1}{c}{\bf Configuration} &  \multicolumn{1}{c}{\bf Beam} & \multicolumn{1}{c}{\bf Resolution (\kms)} & \multicolumn{1}{c}{\bf (mJy beam$^{-1}$)} \\ \hline \hline
 HC$_3$N J = 4--3   & 36.392992 & EW367 & $3.0 \times 16.4$ & 8.2 & 0.4\\
                                 &                  & EW367+1.5A & $0.8 \times 3.1$ & 20.0 & 0.5 \\ 
 H56$\alpha$            & 36.46626  & EW367 & $3.0 \times 16.6$ & 8.2 & 0.4\\
 CH$_3$CN J = 2--1 & 36.795568  & EW367 & $3.0 \times 16.4$ & 8.2 & 0.4\\
 H53$\alpha$            & 42.95197  & EW367 & $2.0 \times 12.0$ & 6.8 & 0.5 \\ 
                                 &                  & EW367+1.5A & $0.7 \times 3.1$ & 20.0 & 0.6 \\ 
 SiO J = 1--0 $\nu$ =0 & 43.423853  & EW367 & $2.7 \times 11.9$ & 6.8 & 0.5\\
                                  &                  & EW367+1.5A & $0.7 \times 3.1$ & 20.0 & 0.7\\ 
 HNCO (2$_{0,2}$-1$_0,1$) & 43.962998  & EW367 & $2.6 \times 11.8$ & 8.2 & 0.5\\
                                  &                  & EW367+1.5A & $0.7 \times 3.1$ & 20.0 & 0.7\\ 
 \hline \hline
\end{tabular} \label{tab:molecules}
\end{table*}

\begin{figure*}
   \includegraphics[scale=0.30]{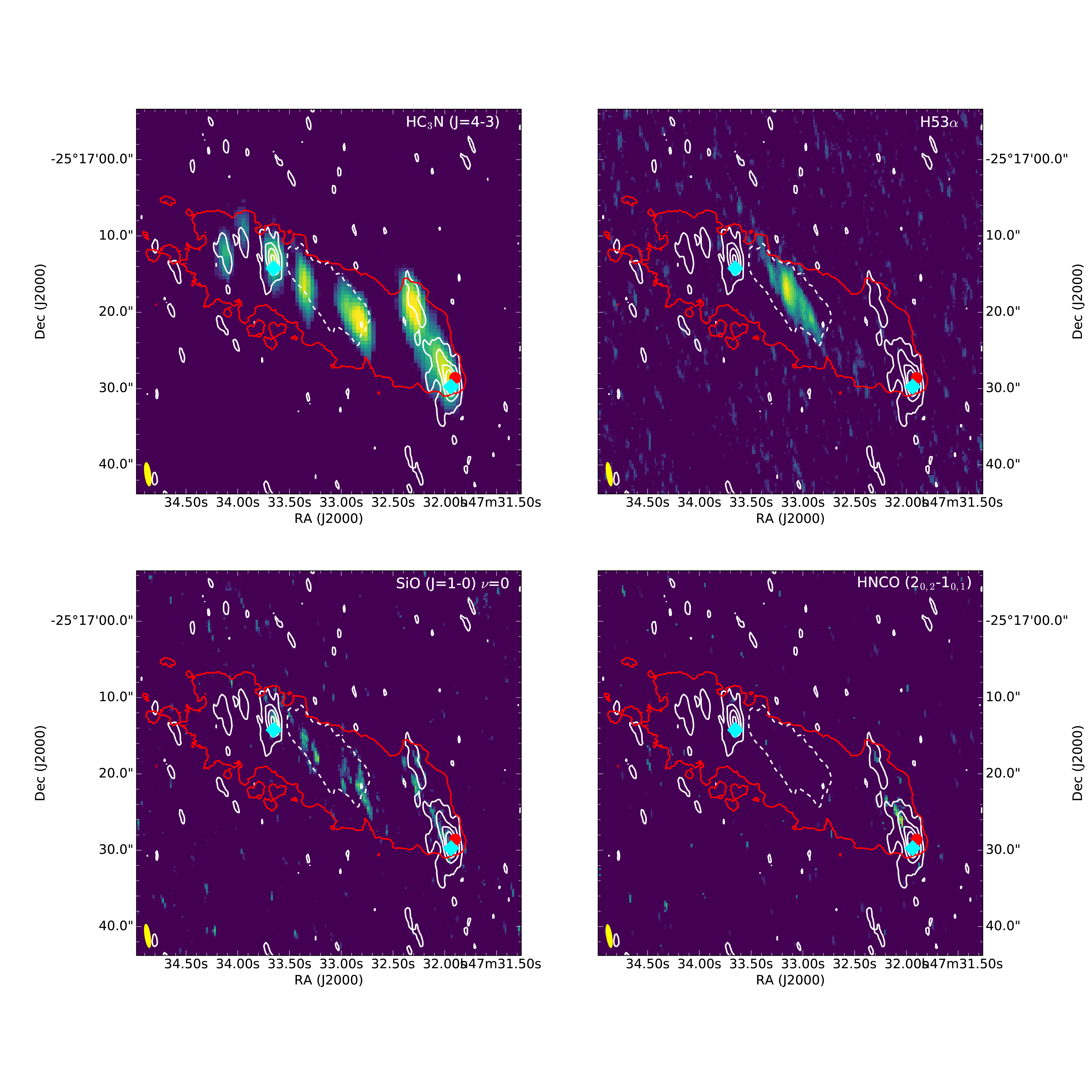}
   \caption{Molecular and atomic emission from transitions observed simultaneously with the observations of the 36.2- and 44.1-GHz methanol transitions imaged using the combined intermediate- and high-resolution data.  The other details are as for Figure~\ref{fig:molecules}.}
  \label{fig:moremolecules}
\end{figure*}

\section{Discussion}

Figure~\ref{fig:ngc253} shows the 36.2-GHz methanol emission with respect to the continuum emission, other molecular transitions and objects of interest.  The methanol emission is associated with molecular cores observed in CO \citep{Sakamoto+06,Sakamoto+11,Leroy+15}, \ammonia\/ \citep{Ott+05,Lebron+11,Gorski+17}, CS, HCN and HCO+ \citep{Knudsen+07,Leroy+15,Meier+15} and offset from the location of supershells \citep{Sakamoto+06,Bolatto+13}.  \cite{Leroy+15} used multiple ALMA configurations to investigate both the CO (1-0) emission and a number of 3-mm dense gas tracers.  From these data they identified 10 giant molecular clouds (GMC) in the nuclear region of NGC\,253 and these are marked as yellow squares in Figure~\ref{fig:ngc253}.  Four of the ten GMCs \citep[][sources 1, 7, 8 and 9]{Leroy+15} are associated with the methanol masers and the offsets of the different methanol maser components from the clouds are listed in Table~\ref{tab:component}.

Recent Jansky Very Large Array (JVLA) observations of NGC\,253 by \citet{Gorski+17} covered a number of molecular lines in the frequency range 21--37~GHz, including the 36.2-GHz methanol transition.  Those observations have similar angular resolution and sensitivity to those presented here and show similar results, with offsets between the measured positions of the methanol emission of around 0.5 arcseconds or less (see Table~\ref{tab:component}).  Comparison of the three observations of the 36.2-GHz methanol undertaken with the ATCA (Section~\ref{sec:results}) showed the H168 observations made in 2014 March had significantly lower integrated flux density than the higher resolution observations from 2014 October and November.  The JVLA observations of \citet{Gorski+17} were made in 2013 May and show the peak of the integrated intensity from the MM1-MM3 region to be around 40 mJy km/s, which lies between the intensities seen in our intermediate- and high-resolution observations.  The spectra shown in figure~8 of \citet{Gorski+17} were extracted from data with angular resolution 6 $\times$ 4 arcseconds, which is between between the angular resolution of the intermediate- and high-resolution observations (Table~\ref{tab:obs}) and so is consistent with some of the 36.2-GHz methanol emission becoming resolved out at higher angular resolution.  The luminosities calculated from the integrated intensity of the methanol emission from the JVLA observations are also in reasonable agreement with those obtained from the current ATCA observations (see Section~\ref{sec:masers}).

\subsection{Comparison of methanol with other molecules} \label{sec:comparison}

Many different molecular transitions have been observed at high resolution and sensitivity toward NGC\,253 \citep[e.g.][]{Leroy+15,Meier+15,Gorski+17} and we can compare these with the 36.2- and 44.1-GHz methanol emission to obtain insights into their relationship with broader galactic morphology.  The high-resolution observations of the thermal molecular and atomic transitions resolve much of the emission (as expected).   Figure~\ref{fig:moremolecules} shows images using combined data from the two array configurations for all the molecular transitions except CH$_3$CN J = 2--1, which was the weakest of the transitions detected and is not detected in the combined array image.  These images yield information on any compact components of the molecular and atomic transitions.  Perhaps the most striking aspect of these images is the close correlation between the \cyanoacet\/ J = 4--3 and 36.2-GHz methanol maser emission.  Compact \cyanoacet\/ emission is observed towards all the regions which show compact 36.2-GHz methanol emission.  This was investigated in more detail by \citet{Ellingsen+17} who found that the MM1 region also hosts a \cyanoacet\/ J = 4--3 maser, detected at 0.1 arcsecond resolution in JVLA data.  \cyanoacet\/ also shows compact emission from two regions towards the outer edges of the central continuum emission where no 36.2-GHz methanol is detected.  The other two molecular transitions are nearly entirely resolved out at the resolution of the combined array images.  Regions of weak, compact SiO emission are observed, mainly towards the central continuum source, but some towards the western MM1-MM3 methanol complex.  The HNCO emission shows weak emission close to the MM1-MM2 methanol sources, but essentially nowhere else.

\citet{Leroy+15} and \citet{Meier+15} used ALMA to make observations of a large number of molecular transitions in the frequency 85.6 -- 115.7~GHz (covering approximately 50 percent of this frequency range) with angular resolution in the range 2--4 arcseconds.  \citeauthor{Leroy+15} focus on the dense gas tracers and comparison of those with CO and its isotopologues, while \citeauthor{Meier+15} cover the remaining molecular transitions and make detailed comparison of the distribution of dense gas, shock tracers, photo dissociation regions (PDR), recombination lines and total molecular gas. There are a number of molecules where \citeauthor{Meier+15} observed a higher excitation transition of a molecule covered in the current observations, specifically \cyanoacet\/ J = 11--10, SiO J = 2--1 $\nu$ =0 and HNCO ($4_{0,4} - 3_{0,3}$).  For each of these the overall morphology seen in the ALMA observations is very similar to that observed at lower angular resolution by the ATCA (Figure~\ref{fig:molecules}).  From LVG modelling of a number of thermal transitions they find that the bulk of the molecular gas in the nuclear region of NGC\,253 is consistent with kinetic temperatures $T_{kin}$ of around 80~K and densities $n_{H_2}$ of around 10$^{4.75}$ cm$^{-3}$.  \citet{Meier+15} show that PDR tracers are most prominent in the inner portion of the nuclear molecular gas, where they state that star formation is most intense and no 36.2-GHz methanol emission is observed.  The most interesting result in terms of understanding the 36.2-GHz methanol emission is in the relative distribution of the two shock tracers SiO and HNCO.  The SiO emission has a very similar distribution to the optically thin dense gas tracer H$^{13}$CN (see \citet{Meier+15} figure~3), whereas the HNCO emission is relatively enhanced in the outer regions of the nuclear molecular gas compared to the inner regions (similar to the 36.2-GHz methanol emission).  The interpretation of this result is that the in the inner regions HNCO is absent because of a combination of photodissociation and the details of the excitation mechanisms relative to SiO.  Similar enhancement of HNCO relative to SiO is seen in the bar-shock regions of a number of other nearby galaxies \citep{Meier+05,Meier+12}.

\citet{Gorski+17} used observations of six different \ammonia\/ inversion transitions to estimate the kinetic temperature of the gas in the central region of NGC\,253.  They found that their ammonia data required the presence of two different gas components, a cool component with a temperature around 57~K and a warmer gas component with a temperature of 130~K.  Interestingly, there was relatively little variation in temperature observed across the nuclear region and \citet{Gorski+17} interpret this as indicating that the dominant heating is not related to PDRs or shocks (since tracers of these do vary significantly), but that perhaps cosmic rays may be the dominant source of heating.  The \ammonia\/ clouds A1, A5 and A6 are most closely associated with the 36.2-GHz methanol emission regions and while these show temperatures similar to the regions closer to the nucleus they do show significantly larger velocity dispersion.  \citet{Gorski+17} estimate that the mechanical energy in the NGC\,253 molecular gas exceeds the thermal energy by more than two orders of magnitude and in the molecular gas in the outer regions this ratio will be even higher.  This suggests that turbulence may play a significant role in heating the molecular gas in the methanol maser emission regions. 

\subsection{Evidence for maser emission from 36.2- and 44.1-GHz methanol transitions} \label{sec:masers}

\citet{Ellingsen+14} claimed that the 36.2-GHz methanol emission in NGC\,253 was a maser from comparison of emission from the same transition in Galactic sources and comparison with single dish observations of other methanol transitions in NGC\,253.  With higher resolution observations of the methanol emission and additional high resolution data on thermal molecular emission in NGC\,253 it is worth revisiting the evidence as to whether or not the 36.2-GHz emission is a maser.  Unfortunately, there are no published images of other methanol transitions toward NGC\,253 for direct comparison.  In particular, it would be interesting to be able to compare the 36.2-GHz methanol distribution to low-excitation methanol lines such as the 96-GHz $2_k \rightarrow 1_k$ series \citep{Henkel+87} or the 48.3-GHz $1_0 \rightarrow 0_0$A$^+$ and  $1_0 \rightarrow 0_0$E transitions.  

Figure~\ref{fig:ngc253} shows that the brightest 36.2-GHz methanol emission from the combined array data is observed towards the MM1 and MM6 regions and at a velocity resolution of 3~\kms\/ it has a peak intensity of around 12 mJy beam$^{-1}$, corresponding to a brightness temperature of approximately 5~K.  If we instead consider the integrated emission observed in a single synthesised beam, the MM2 region has 0.5 Jy~\kms\/, corresponding to a brightness temperature of 220~K integrated over the 60~\kms\/ velocity range of the emission at this site.  These values are in the range of the rotational and kinetic temperatures observed for molecular emission in the central region of NGC\,253 \citep[e.g.][]{Martin+06,Meier+15} and so do not in themselves imply maser emission.  The \cyanoacet\/ J = 4--3 transition shows the most similar spatial distribution to the 36.2-GHz methanol in the combined array images (Figure~\ref{fig:moremolecules}) and making direct comparison at each of MM1-MM7 we find the two transitions have a very similar velocity range and the intensity of the \cyanoacet\/ transition is about one third that of the methanol (although it does vary significantly between sites).  The complicating factor here is that a  \cyanoacet\/ maser emission has been detected in the MM1 region \citep{Ellingsen+17} and it may be that much of both the \cyanoacet\/ and 36.2-GHz methanol emission detected in the combined array image arises from weak diffuse maser regions with scale sizes in the 10-100~pc range, similar to that observed for extragalactic H$_2$CO masers \citep{Baan+17}. The only other two molecular transitions detected in combined array images were the vibrational ground-state of SiO and HNCO, both of which show only very weak emission from small regions.  This indicates that a much larger fraction of the emission from these transitions is resolved out on angular scales of a few arcseconds.  So the methanol emission shows a very different distribution to thermal emission from molecules other than \cyanoacet\/ and the similarity with \cyanoacet\/ may be because it also exhibits significant weak, diffuse maser emission.

The best studied Galactic class~I methanol masers are associated with high-mass star formation regions \citep[e.g.][]{Plambeck+90,Voronkov+06,Voronkov+14}.  One potential mechanism for producing intense extragalactic class~I methanol maser emission in a starburst system is through combined emission from a large number of normal star formation systems within a small volume of space.  If we take 500~Jy~\kms\/kpc$^2$ ($1.5 \times 10^{-6}$ \lsol) as a typical isotropic luminosity for a Galactic class~I methanol maser, then the total luminosity of the 36.2-GHz methanol emission toward NGC\,253 measured from the intermediate-resolution observations is $6.7 \times 10^{7}$~Jy~\kms\/kpc$^2$ (0.2 \lsol), i.e. more than 100000 times that of a typical Galactic class~I maser.  Considering the strongest 36.2-GHz methanol emission region (MM1/MM2) we have an integrated flux density of 0.5 Jy~\kms\/ within a synthesised beam of $0.9 \times 3.4$ arcseconds, corresponding to linear scales of approximately $60 \times 15$ pc.  This is more than 10000 times the luminosity of a typical Galactic class~I methanol maser from a relatively small volume, implying an average volume of $\sim$ less than 2.5 pc$^{3}$ per high-mass star formation region, which although implausible is not impossible. Within Galactic star formation regions the 36.2-GHz class~I transition is usually accompanied by emission of comparable or greater intensity from the 44.1-GHz class~I transition, as well as 6.7-GHz class~II masers, 22-GHz water masers and other transitions \citep[e.g.][]{Voronkov+14}.  Hence, if the 36.2-GHz methanol emission were due to a large number of Galactic-like high-mass star formation regions then we would expect it to be accompanied by similar (or stronger) masers from a variety of other transitions, but that is not the case, with the 44.1-GHz class~I emission being two orders of magnitude weaker (see section~\ref{sec:results}), no 6.7-GHz class~II methanol masers with peak intensity greater than 110~mJy \citep{Phillips+98a} and no water maser emission with peak flux density greater than a few mJy associated with any of the 36.2-GHz methanol emission regions \citep{Gorski+17}. 

In the scaled-up Galactic star formation region scenario the methanol masers would be associated with regions of young, very intense star formation within the starburst. \citet{Matsubayashi+09} made sensitive observations of the H$\alpha$, S$\mbox{\sc{ii}}$ and N$\mbox{\sc{ii}}$ optical lines and used the H$\alpha$:S$\mbox{\sc{ii}}$ ratio to determine whether star formation or the galactic wind dominates in different regions.  Both the NE and SW methanol emission regions lie beyond the region of intense H$\alpha$ emission, with the NE near the edge of the emission and no optical line emission evident from the SW region.  The absence of any H$\alpha$ emission from the SW region may be due to variable extinction.  However, when this is considered together with the location of the methanol with respect to other molecular species and the findings of \citet{Martin+06}, (that the chemistry of NGC\,253 is consistent with the dominant factor in the nuclear region being low-velocity shocks rather than hot-cores), it is clear that the 36.2-GHz methanol emission cannot be due to a scaled-up Galactic star formation region-like environment.

The 44.1-GHz methanol emission in NGC\,253 is restricted to two very narrow ($<$ 3~\kms) components close to the MM1 and MM4 36.2-GHz regions.  Given what we know about the physical conditions (temperatures and densities) in the molecular gas in the central regions of NGC\,253, the only mechanism capable of producing such narrow lines in a transition with an upper-state energy 64~K above the ground-state, is a maser.  The presence of 44.1-GHz class~I methanol maser emission strengthens the case for some (perhaps most) of the 36.2-GHz methanol emission being a maser and in the discussion below we will assume that this is case.  However, detection of 36.1-GHz methanol emission at angular resolutions an order of magnitude or more better than we have here are required to show this definitively.

\subsection{Methanol masers from cloud-cloud collisions} \label{sec:cloud}
Most previous observations of methanol emission towards NGC\,253 have been made with single dish telescopes pointed at the nuclear region of the galaxy.  The first detection of extragalactic methanol was made towards NGC\,253 by \citet{Henkel+87} who detected the 96-GHz 2$_{k} \rightarrow 1_{k}$ series with the IRAM 30-m telescope with a 24 arcsecond beam.  They detected two spectral components in the methanol emission with peak velocities of around 180 and 300~\kms.  From observations made at a number of offset locations they were able to make a coarse image of the methanol emission and their data suggests that it peaks slightly to the south-west of their pointing centre ($\alpha = 00^{\mbox{h}}47^{\mbox{m}}33.40^{\mbox{s}}$ ; $\delta = -25^\circ 17\arcmin14.0\arcsec$ (J2000)). Note that all the velocities in the current paper are given in the Barycentric reference frame, which for NGC\,253 are 7.24~\kms\/ higher than the equivalent velocity in the local-standard of rest (LSR) frame.

The most detailed study of methanol emission towards NGC\,253 was made as part of the 2-mm spectral line scan of the galaxy by \citet{Martin+06}.  Those observations were made with the IRAM 30-m telescope using a single pointing towards  $\alpha = 00^{\mbox{h}}47^{\mbox{m}}33.3^{\mbox{s}}$ ; $\delta = -25^\circ 17\arcmin23.0\arcsec$ (J2000) (offset 5.8 arcseconds southeast from the pointing centre of the ATCA observations ; see Figure~\ref{fig:ngc253}).  The beamwidth (to half-power) of the IRAM 30-m was in the range 14--19 arcseconds at the two extremes of the observed frequency range (129 and 175~GHz).  The consequence of this is that the observations of \citeauthor{Martin+06} include the very centre of NGC\,253 but not the region where we detected 36.2~GHz methanol.  \citet{Martin+06} detected 14 different methanol transitions in their 129--175~GHz spectral scan, essentially representing 9 independent measurements (as a number of the transitions are in blended series).  From these observations they obtain a rotational temperature of 12~K for the methanol emission and the 36.2-GHz methanol emission observed towards NGC\,253 is a factor of more than 30 stronger than would be predicted for thermal emission from this transition \citep{Ellingsen+14}.

For some of the methanol transitions \citet{Martin+06} report multiple spectral components with Barycentric velocities of around 170 and 300~\kms, similar to the velocities of the northeast and southwest components of the 36.2~GHz methanol emission. The frequency coverage of their spectral scan includes 6 different known methanol maser transitions.  Two of these are class~I maser transitions, specifically the 132.9-GHz ($6_{-1} \rightarrow 5_{0}E$) transition is in the same family as the 36.2-GHz maser and the 146.6-GHz ($9_{0} \rightarrow 8_{-1}A^+$) transition is in the same family as the 44.1-GHz maser.  Neither of these class~I transitions were detected in the \citet{Martin+06} spectral scan.  The other four known maser transitions are all class~II transitions, at 148.1 GHz ($15_{0} \rightarrow 15_{-1}E$), 156.2 GHz ($2_{1} \rightarrow 3_{0}A^{+}$) 157.3 GHz ($J_{0} \rightarrow J_{-1}E$) and 165.0 GHz ($J_{1} \rightarrow J_{0}E$).  The 157.3 and 165.0-GHz transition series were both detected by \citet{Martin+05}, but there is no suggestion from their data of significant non-thermal behaviour.  However, in the absence of images with comparable angular resolution of any established thermal methanol emission, we do not know of the relative importance of the distribution of methanol compared to the excitation conditions in the absence of 36.2-GHz methanol emission in the region of NGC\,253 covered by the primary beam of the \citeauthor{Martin+06} observations.

As discussed in Section~\ref{sec:comparison}, the spatial distribution of the methanol emission differs from that of dense gas tracers such as \ammonia, CS, HCN and HCO$^{+}$ \citep{Ott+05,Knudsen+07,Leroy+15,Meier+15}, which show emission throughout the majority of the nuclear region. The 36.2-GHz methanol is located at the extreme ends of the molecular emission from other species.  \citet{Garcia-Burillo+00} imaged the thermal SiO J = 2--1 $\nu$ =0 emission towards NGC\,253 and showed that it is concentrated in four regions, two either side of the kinematic centre.  They suggest that these four regions are associated with the Inner Linblad Resonances (ILR).  The more intense SiO emission closer to the nucleus being associated with the inner ILR (iIlR) and the weaker, more diffuse regions being associated with the outer ILR (oILR).  The dynamics of the bar lead to a concentration of material at the ILR and \citeauthor{Garcia-Burillo+00} suggest that the SiO is produced by large-scale collisions in these regions.  Figure~\ref{fig:molecules} shows the integrated intensity of the SiO J = 1--0 $\nu$ =0 transition with the integrated 36.2-GHz methanol emission superimposed.  This lower-J transition exhibits a similar spatial distribution to that shown by \citet{Garcia-Burillo+00}.  The 36.2-GHz methanol masers are associated with the outer SiO emission regions, and are marginally offset towards greater radii.  This suggests that the methanol maser emission arises at the oILR, where there are significant cloud-cloud collisions at the interface between the nuclear ring and the inner end of the galactic bar.  Additional support for this hypothesis comes from \citet{Iodice+14} who use near-infrared photometry to investigate the dynamics of NGC\,253 and estimate the location of the ILR to be between 300 and 400~pc from the nucleus, consistent with the south-western 36.2-GHz methanol emission.

The relative chemical abundance of different molecular species can be used to infer the age of starburst systems and the current dominant physical processes responsible for the emission. \citet{Martin+05,Martin+06} found that the abundance of molecules such as CH$_3$CN, HC$_3$N, methanol, formaldehyde and OCS in the central region of NGC\,253 are very similar to Galactic centre molecular clouds where the chemistry is dominated by low velocity shocks.  The high HNCO:SiO ratio and larger velocity dispersion of the outer nuclear region molecular clouds (discussed in Section~\ref{sec:comparison}) are also consistent with this interpretation. Furthermore, widespread strong 36.2-GHz methanol maser emission is detected towards the centre of the Milky Way by \citet{Yusef-Zadeh+13}, accompanied by much weaker 44.1-GHz methanol masers \citep{McEwen+16}.  If 36.2-GHz emission similar to that detected towards the central region of the Milky Way were present towards the centre of NGC\,253 it would have an integrated intensity of around 35 mJy~\kms.  This is less than 1 percent of the integrated intensity detected in the intermediate resolution data, however, it would have been readily detectable at the sensitivity of those observations.   The absence of 36.2-GHz methanol emission in the central region of NGC\,253 may be due to a lower abundance of methanol in this region because of dissociation from high energy photons or cosmic rays, alternatively it may be the absence of large-scale low-velocity shocks capable of producing class~I methanol masers.

The other Galactic environment where class~I methanol maser emission has been studied in some detail is in regions where supernova remnants are interacting with molecular clouds.  In those regions the 36.2-GHz methanol masers are often stronger than the 44.1-GHz masers and sometimes the latter are completely absent \cite[e.g.][]{Pihlstrom+14,McEwen+16}.  Modelling of class~I methanol masers by \citet{McEwen+14} and \citet{Leurini+16} shows that this occurs for gas at relatively higher densities ($n_{H_2} > 10^{7}$ cm$^{-3}$) and temperatures ($T \ge 100$ K) than the conditions under which the 44.1-GHz maser is the strongest class~I methanol transition.  \citet{Gorski+17} show that much of the 36.2-GHz methanol emission is located at the edge of supershells observed in CO and dense gas tracers \citep{Sakamoto+11,Bolatto+13} and speculate that it could be produced by a large number of protostellar outflows or supernova remnants in these regions.  In Section~\ref{sec:masers} we demonstrated the significant differences between the 36.2 and 44.1-GHz methanol emission observed in NGC\,253 and that observed in typical Galactic high-mass star formation regions \cite[e.g.][]{Voronkov+14}, which would seem to rule out individual protostellar outflows as the mechanism.  Supernovae in NGC\,253 have also been studied in some detail \citep[e.g.][]{Lenc+06}, but they are restricted to the inner nuclear region which shows no 36.2-GHz methanol emission.  

We suggest that scenario for the production of 36.2-GHz methanol masers which best fits the observational data for NGC\,253 is that they arise in regions where there are large-scale, low-velocity shocks, but a lower radiation field than the inner nuclear zone (where we observe continuum emission, radio recombination lines and molecular transitions associated with PDRs).  The strong south-western 36.2-GHz methanol masers can be explained as being produced in a region where molecular gas is being compressed and shocked by the combined effect of expansion of a superbubble and inflowing molecular material from the galactic bar.  The north-eastern 36.2-GHz methanol masers are generally weaker, but here the methanol partially overlaps with emission from radio recombination lines (e.g. H56$\alpha$), suggesting a higher radiation field, which is also consistent with weaker emission seen in other molecules sensitive to photodissociation, such as HNCO.  So it may be that in the north-eastern region there is more active or advanced star formation, or other energetic processes than the south-western region and the resulting radiation is in the process of destroying the gas-phase methanol in this region.  The methanol maser emission in NGC\,253 is thus envisaged as a scaled-up version of that seen in the central regions of the Milky Way and in supernova-molecular cloud interactions, perhaps with a substantial fraction of the emission arising in larger-scale, diffuse regions, rather than compact components.  Regions of weak, diffuse maser emission do not appear to be consistent with the current model requirements for 36.2-GHz masers to dominate the 44.1-GHz transition at higher densities, however, recent work in this area has focused on compact maser emission, so there may also be regimes at lower densities where 36.2-GHz masers are stronger which have yet to be identified.

If the conditions which produce class~I methanol masers in NGC\,253 are broadly speaking, a large-scale form of supernova-molecular cloud interactions, then we might predict that other shock-excited masers, such as the 1720-MHz OH maser emission may also be present close to the class~I methanol masers \citep{Frail11}.  The ground-state OH lines in NGC\,253 were imaged with the VLA by \citet{Turner+85} at $17.7 \times 9.5$ arcsecond resolution and used to investigate the gas-dynamics in the nuclear region.  At these angular scales the OH spectra are dominated absorption in the foreground of the nuclear region.  Interestingly, \citet{Turner+85} comment that in the 1720-MHz OH transition they observe strong emission west of the nucleus dominating the absorption which is not seen in other transitions, nor amplifying the background continuum emission.  This suggests that further investigations with higher angular resolution (to resolve out foreground absorption) may be warranted.

M82 is the other proto-typical nearby nuclear starburst system, but it has a much lower abundance of methanol than NGC\,253 \citep{Martin+06b}.  It is thought that M82 is a more evolved starburst with the chemistry dominated by PDRs, in contrast to the intermediate age starburst of NGC\,253 \citep{Martin+05}.  M82 has not yet been searched for 36.2-GHz methanol maser emission, however, we might expect that class~I methanol maser emission from the nuclear region of M82 will be absent, or significantly weaker than is the case for NGC\,253.  \citet{Martin+06} compared the chemistry of NGC\,253 with that of five other nearby starburst galaxies M82, IC\,342, Maffei 2, NGC\,4945 and NGC\,6946.  They found NGC\,4945 and IC\,342 to be the most chemically similar to NGC\,253, which suggests that they are both good potential targets for future searches for extragalactic 36.2-GHz methanol emission, particularly as a number of these sources show similar enhancement in HNCO:SiO ratios due to bar shocks \citep{Meier+12}.

\subsection{Methanol as a probe of fundamental constants} \label{sec:constants}
The methanol molecule has a rich rotational spectrum due to the hindered internal rotation of the OH radical about the C--O bond.  This hindered internal rotation also makes the rest frequencies of the rotational transitions of methanol unusually sensitive to the proton-to-electron mass ratio \citep{Levshakov+11,Jansen+11a}.  Comparison of the rest frequency observed for a spectral line astronomically with that observed in the laboratory can be used to test for variations in the fundamental constants of physics, such as the ratio of the mass of the proton and electron or the ratio of Planck's constant to the speed of light \citep[e.g.][]{Bagdonaite+13,Kanekar11}.  The observed frequency of a spectral line in an astronomical observation depends upon the relative motion of the source along the line of sight to the observer.  To determine if the rest frequency of a line has changed compared to the expectations from the laboratory we have to compare two different spectral lines which have different dependencies on the fundamental constants of physics.  However, in general different spectral lines will trace different physical conditions and hence environments, which introduces additional uncertainties in the relative line of sight velocities \citep[see][for more detailed discussion]{Ellingsen+11b}. 

The different rotational transitions of methanol show significant differences in their sensitivity to the proton-to-electron mass ratio $\mu$ \citep{Levshakov+11,Jansen+11} and some Galactic methanol masers have been demonstrated to have the emission from the 6.7 and 12.2-GHz transitions co-spatial at the milli-arcsecond scale \citep{Menten+92}, as well as exhibiting both narrow and intense emission.  The 6.7- and 12.2-GHz class II methanol maser transitions are the strongest and best studied Galactic methanol maser transitions.  Unfortunately, previous attempts to detect class~II methanol masers in extragalactic sources have been largely unsuccessful, with only a handful of sources detected in local group galaxies \citep[e.g.][]{Green+08,Ellingsen+10,Sjouwerman+10}.  Despite extensive searches there have been no class~II methanol megamasers detected \citep{Ellingsen+94,Phillips+98b,Darling+03}.  The extragalactic class~II methanol masers detected to date are less luminous than the strongest Galactic sources and so there appears little prospect of being able to detect them at cosmologically interesting distances.  The recent detection of luminous class~I methanol masers towards two nearby starburst galaxies \citep{Ellingsen+14,Wang+14} and the major merger system Arp\,220 \citep{Chen+15}, combined with finding by \citet{Chen+16} of a possible relationship between methanol maser luminosity and star formation rate, suggests that it may be possible to detect methanol megamasers at cosmologically interesting distances.

Observations of multiple methanol maser transitions towards the same source then opens the prospect of a new and sensitive probe for changes in the proton-to-electron mass ratio. The most sensitive constraints on $\mu$ to date have been achieved through observations of absorption from different methanol transitions towards the lensing galaxy in the PKS\,B1830-211 gravitational lens system \citep{Ellingsen+12b,Bagdonaite+13}. Strong lensing systems such as PKS\,B1830-211 are rare, however, major mergers are relatively common, so there is the potential to undertake a sensitive test for changes in the proton-to-electron mass ratio over a range of redshifts distributed across the sky.  As methanol contains both a carbon and oxygen atom we expect its abundance to be significantly reduced in low metallicity systems, hence, the degree to which methanol can be used to test for changes in the fundamental constants at high redshift will depend on the degree to which metallicity impacts the luminosity of class~I methanol masers in starburst systems \citep[e.g.][]{Breen+11,Ellingsen+10}.  The fundamental limitation will then be the degree to which the emission from different methanol maser transitions is co-spatial and the relative sensitivity to changes in $\mu$ of the different transitions.  For the 36.2-, 44.1- and 84.5-GHz class~I transitions (the transitions detected in extragalactic sources to date), the sensitivities are 9.6, 5.3 and 3.5, respectively \citep{Jansen+11,Levshakov+11}.  For NGC\,253 we have found the 44.1-GHz methanol emission to be significantly weaker than the 36.2-GHz transition and to have very different spectral profiles.  This suggests that these two transitions are probably not suitable for sensitive tests of the proton-to-electron mass ratio and that it will be necessary to identify another methanol transition which is more similar to the 36.2-GHz to make this an interesting prospect.

\section{Conclusions}

We have made sensitive arcsecond-resolution images of the 36.2- and 44.1-GHz class~I methanol transitions towards the nuclear region of NGC\,253.  The methanol emission is located towards the edges of the nuclear molecular gas in regions where there is ionising radiation to dissociate the molecules.  We have detected the first extragalactic 44.1-GHz methanol masers, which are associated with regions of stronger 36.2-GHz methanol masers.   The class~I masers are located at the inner end of the galactic bar, close to the inner Linblad resonance and are likely due to large-scale cloud-cloud collisions occurring in this region.  The characteristics of the NGC\,253 methanol masers appear to be similar to those of the methanol masers detected towards the central region of the Milky Way and in supernova-molecular cloud interactions.  We suggest that nearby spiral galaxies with bar shocks are be prime targets to search for other extragalactic class~I methanol masers to better characterise and understand their relationship with galactic morphology and dynamics.

\section*{Acknowledgements}

The ATCA is part of the Australia Telescope which is funded by the Commonwealth of Australia for operation as a National Facility managed by CSIRO.  We thank the referee Dr Anita Richards, for constructive comments which have improved the manuscript.  X.C. acknowledges support from National Natural Science Foundation of China (NSFC) through grant NSFC 11590781. This research made use of NASA's Astrophysics Data System Abstract Service. This research has made use of the NASA/IPAC Extragalactic Database (NED) which is operated by the Jet Propulsion Laboratory, California Institute of Technology, under contract with the National Aeronautics and Space Administration.

\bibliography{references}

\begin{thebibliography}{68}
\expandafter\ifx\csname natexlab\endcsname\relax\def\natexlab#1{#1}\fi

\bibitem[{{Baan} {et~al}\mbox{.}(2017){Baan}, {An}, {Kl{\"o}ckner}, \&
  {Thomasson}}]{Baan+17}
{Baan} W.~A., {An} T., {Kl{\"o}ckner} H.-R., {Thomasson} P., 2017, \mnras, 469,
  916

\bibitem[{{Bagdonaite} {et~al}\mbox{.}(2013){Bagdonaite}, {Jansen}, {Henkel},
  {Bethlem}, {Menten}, \& {Ubachs}}]{Bagdonaite+13}
{Bagdonaite} J., {Jansen} P., {Henkel} C., {Bethlem} H.~L., {Menten} K.~M.,
  {Ubachs} W., 2013, Science, 339, 46

\bibitem[{{Bolatto} {et~al}\mbox{.}(2013){Bolatto}, {Warren}, {Leroy},
  {Walter}, {Veilleux}, {Ostriker}, {Ott}, {Zwaan}, {Fisher}, {Weiss},
  {Rosolowsky}, \& {Hodge}}]{Bolatto+13}
{Bolatto} A.~D. {et~al.}, 2013, \nat, 499, 450

\bibitem[{{Breen} {et~al}\mbox{.}(2011){Breen}, {Ellingsen}, {Caswell},
  {Green}, {Fuller}, {Voronkov}, {Quinn}, \& {Avison}}]{Breen+11}
{Breen} S.~L., {Ellingsen} S.~P., {Caswell} J.~L., {Green} J.~A., {Fuller}
  G.~A., {Voronkov} M.~A., {Quinn} L.~J., {Avison} A., 2011, \apj, 733, 80

\bibitem[{{Brunthaler} {et~al}\mbox{.}(2009){Brunthaler}, {Castangia},
  {Tarchi}, {Henkel}, {Reid}, {Falcke}, \& {Menten}}]{Brunthaler+09}
{Brunthaler} A., {Castangia} P., {Tarchi} A., {Henkel} C., {Reid} M.~J.,
  {Falcke} H., {Menten} K.~M., 2009, \aap, 497, 103

\bibitem[{{Chen} {et~al}\mbox{.}(2015){Chen}, {Ellingsen}, {Baan}, {Qiao},
  {Li}, {An}, \& {Breen}}]{Chen+15}
{Chen} X., {Ellingsen} S.~P., {Baan} W.~A., {Qiao} H.-H., {Li} J., {An} T.,
  {Breen} S.~L., 2015, \apjl, 800, L2

\bibitem[{{Chen} {et~al}\mbox{.}(2016){Chen}, {Ellingsen}, {Zhang}, {Wang},
  {Shen}, {Wu}, \& {Wu}}]{Chen+16}
{Chen} X., {Ellingsen} S.~P., {Zhang} J.-S., {Wang} J.-Z., {Shen} Z.-Q., {Wu}
  Q.-W., {Wu} Z.-Z., 2016, \mnras, 459, 357

\bibitem[{{Dalcanton} {et~al}\mbox{.}(2009){Dalcanton}, {Williams}, {Seth},
  {Dolphin}, {Holtzman}, {Rosema}, {Skillman}, {Cole}, {Girardi}, {Gogarten},
  {Karachentsev}, {Olsen}, {Weisz}, {Christensen}, {Freeman}, {Gilbert},
  {Gallart}, {Harris}, {Hodge}, {de Jong}, {Karachentseva}, {Mateo}, {Stetson},
  {Tavarez}, {Zaritsky}, {Governato}, \& {Quinn}}]{Dalcanton+09}
{Dalcanton} J.~J. {et~al.}, 2009, \apjs, 183, 67

\bibitem[{{Dale} {et~al}\mbox{.}(2009){Dale}, {Cohen}, {Johnson}, {Schuster},
  {Calzetti}, {Engelbracht}, {Gil de Paz}, {Kennicutt}, {Lee}, {Begum},
  {Block}, {Dalcanton}, {Funes}, {Gordon}, {Johnson}, {Marble}, {Sakai},
  {Skillman}, {van Zee}, {Walter}, {Weisz}, {Williams}, {Wu}, \&
  {Wu}}]{Dale+09}
{Dale} D.~A. {et~al.}, 2009, \apj, 703, 517

\bibitem[{{Darling} {et~al}\mbox{.}(2003){Darling}, {Goldsmith}, {Li}, \&
  {Giovanelli}}]{Darling+03}
{Darling} J., {Goldsmith} P., {Li} D., {Giovanelli} R., 2003, \aj, 125, 1177

\bibitem[{{Deguchi} {et~al}\mbox{.}(2004){Deguchi}, {Fujii}, {Glass}, {Imai},
  {Ita}, {Izumiura}, {Kameya}, {Miyazaki}, {Nakada}, \&
  {Nakashima}}]{Deguchi+04}
{Deguchi} S. {et~al.}, 2004, \pasj, 56, 765

\bibitem[{{Ellingsen} {et~al}\mbox{.}(2017){Ellingsen}, {Chen}, {Breen}, \&
  {Qiao}}]{Ellingsen+17}
{Ellingsen} S., {Chen} X., {Breen} S., {Qiao} H.-H., 2017, \apjl, 841, L14

\bibitem[{{Ellingsen} {et~al}\mbox{.}(2011){Ellingsen}, {Voronkov}, \&
  {Breen}}]{Ellingsen+11b}
{Ellingsen} S., {Voronkov} M., {Breen} S., 2011, Physical Review Letters, 107,
  270801

\bibitem[{{Ellingsen} {et~al}\mbox{.}(2010){Ellingsen}, {Breen}, {Caswell},
  {Quinn}, \& {Fuller}}]{Ellingsen+10}
{Ellingsen} S.~P., {Breen} S.~L., {Caswell} J.~L., {Quinn} L.~J., {Fuller}
  G.~A., 2010, \mnras, 404, 779

\bibitem[{{Ellingsen} {et~al}\mbox{.}(2014){Ellingsen}, {Chen}, {Qiao}, {Baan},
  {An}, {Li}, \& {Breen}}]{Ellingsen+14}
{Ellingsen} S.~P., {Chen} X., {Qiao} H.-H., {Baan} W., {An} T., {Li} J.,
  {Breen} S.~L., 2014, \apjl, 790, L28

\bibitem[{{Ellingsen} {et~al}\mbox{.}(1994){Ellingsen}, {Norris}, {Whiteoak},
  {Vaile}, {McCullch}, \& {Price}}]{Ellingsen+94}
{Ellingsen} S.~P., {Norris} R.~P., {Whiteoak} J.~B., {Vaile} R.~A., {McCullch}
  P.~M., {Price} M.~G., 1994, \mnras, 267, 510

\bibitem[{{Ellingsen} {et~al}\mbox{.}(2012){Ellingsen}, {Voronkov}, {Breen}, \&
  {Lovell}}]{Ellingsen+12b}
{Ellingsen} S.~P., {Voronkov} M.~A., {Breen} S.~L., {Lovell} J.~E.~J., 2012,
  \apjl, 747, L7

\bibitem[{{Frail}(2011)}]{Frail11}
{Frail} D.~A., 2011, \memsai, 82, 703

\bibitem[{{Garc{\'{\i}}a-Burillo} {et~al}\mbox{.}(2000){Garc{\'{\i}}a-Burillo},
  {Mart{\'{\i}}n-Pintado}, {Fuente}, \& {Neri}}]{Garcia-Burillo+00}
{Garc{\'{\i}}a-Burillo} S., {Mart{\'{\i}}n-Pintado} J., {Fuente} A., {Neri} R.,
  2000, \aap, 355, 499

\bibitem[{{Goddi} {et~al}\mbox{.}(2009){Goddi}, {Greenhill}, {Chandler},
  {Humphreys}, {Matthews}, \& {Gray}}]{Goddi+09}
{Goddi} C., {Greenhill} L.~J., {Chandler} C.~J., {Humphreys} E.~M.~L.,
  {Matthews} L.~D., {Gray} M.~D., 2009, \apj, 698, 1165

\bibitem[{{Gorski} {et~al}\mbox{.}(2017){Gorski}, {Ott}, {Rand}, {Meier},
  {Momjian}, \& {Schinnerer}}]{Gorski+17}
{Gorski} M., {Ott} J., {Rand} R., {Meier} D.~S., {Momjian} E., {Schinnerer} E.,
  2017, \apj, 842, 124

\bibitem[{{Green} {et~al}\mbox{.}(2008){Green}, {Caswell}, {Fuller}, {Breen},
  {Brooks}, {Burton}, {Chrysostomou}, {Cox}, {Diamond}, {Ellingsen}, {Gray},
  {Hoare}, {Masheder}, {McClure-Griffiths}, {Pestalozzi}, {Phillips}, {Quinn},
  {Thompson}, {Voronkov}, {Walsh}, {Ward-Thompson}, {Wong-McSweeney}, {Yates},
  \& {Cohen}}]{Green+08}
{Green} J.~A. {et~al.}, 2008, \mnras, 385, 948

\bibitem[{{Henkel} {et~al}\mbox{.}(1987){Henkel}, {Jacq}, {Mauersberger},
  {Menten}, \& {Steppe}}]{Henkel+87}
{Henkel} C., {Jacq} T., {Mauersberger} R., {Menten} K.~M., {Steppe} H., 1987,
  \aap, 188, L1

\bibitem[{{Henkel} {et~al}\mbox{.}(2004){Henkel}, {Tarchi}, {Menten}, \&
  {Peck}}]{Henkel+04}
{Henkel} C., {Tarchi} A., {Menten} K.~M., {Peck} A.~B., 2004, \aap, 414, 117

\bibitem[{{Hofner} {et~al}\mbox{.}(2006){Hofner}, {Baan}, \&
  {Takano}}]{Hofner+06}
{Hofner} P., {Baan} W.~A., {Takano} S., 2006, \aj, 131, 2074

\bibitem[{{Hunt} {et~al}\mbox{.}(1999){Hunt}, {Whiteoak}, {Cragg}, {White}, \&
  {Jones}}]{Hunt+99}
{Hunt} M.~R., {Whiteoak} J.~B., {Cragg} D.~M., {White} G.~L., {Jones} P.~A.,
  1999, \mnras, 302, 1

\bibitem[{{Iodice} {et~al}\mbox{.}(2014){Iodice}, {Arnaboldi}, {Rejkuba},
  {Neeser}, {Greggio}, {Gonzalez}, {Irwin}, \& {Emerson}}]{Iodice+14}
{Iodice} E., {Arnaboldi} M., {Rejkuba} M., {Neeser} M.~J., {Greggio} L.,
  {Gonzalez} O.~A., {Irwin} M., {Emerson} J.~P., 2014, \aap, 567, A86

\bibitem[{{Jansen} {et~al}\mbox{.}(2011{\natexlab{a}}){Jansen}, {Xu},
  {Kleiner}, {Ubachs}, \& {Bethlem}}]{Jansen+11a}
{Jansen} P., {Xu} L.-H., {Kleiner} I., {Ubachs} W., {Bethlem} H.~L.,
  2011{\natexlab{a}}, Physical Review Letters, 106, 100801

\bibitem[{{Jansen} {et~al}\mbox{.}(2011{\natexlab{b}}){Jansen}, {Xu},
  {Kleiner}, {Ubachs}, \& {Bethlem}}]{Jansen+11}
{Jansen} P., {Xu} L.-H., {Kleiner} I., {Ubachs} W., {Bethlem} H.~L.,
  2011{\natexlab{b}}, Physical Review Letters, 106, 100801

\bibitem[{{Kanekar}(2011)}]{Kanekar11}
{Kanekar} N., 2011, \apjl, 728, L12

\bibitem[{{Knudsen} {et~al}\mbox{.}(2007){Knudsen}, {Walter}, {Weiss},
  {Bolatto}, {Riechers}, \& {Menten}}]{Knudsen+07}
{Knudsen} K.~K., {Walter} F., {Weiss} A., {Bolatto} A., {Riechers} D.~A.,
  {Menten} K., 2007, \apj, 666, 156

\bibitem[{{Koribalski} {et~al}\mbox{.}(2004){Koribalski}, {Staveley-Smith},
  {Kilborn}, {Ryder}, {Kraan-Korteweg}, {Ryan-Weber}, {Ekers}, {Jerjen},
  {Henning}, {Putman}, {Zwaan}, {de Blok}, {Calabretta}, {Disney}, {Minchin},
  {Bhathal}, {Boyce}, {Drinkwater}, {Freeman}, {Gibson}, {Green}, {Haynes},
  {Juraszek}, {Kesteven}, {Knezek}, {Mader}, {Marquarding}, {Meyer}, {Mould},
  {Oosterloo}, {O'Brien}, {Price}, {Sadler}, {Schr{\"o}der}, {Stewart},
  {Stootman}, {Waugh}, {Warren}, {Webster}, \& {Wright}}]{Koribalski+04}
{Koribalski} B.~S. {et~al.}, 2004, \aj, 128, 16

\bibitem[{{Lebr{\'o}n} {et~al}\mbox{.}(2011){Lebr{\'o}n}, {Mangum},
  {Mauersberger}, {Henkel}, {Peck}, {Menten}, {Tarchi}, \&
  {Wei{\ss}}}]{Lebron+11}
{Lebr{\'o}n} M., {Mangum} J.~G., {Mauersberger} R., {Henkel} C., {Peck} A.~B.,
  {Menten} K.~M., {Tarchi} A., {Wei{\ss}} A., 2011, \aap, 534, A56

\bibitem[{{Lehmer} {et~al}\mbox{.}(2013){Lehmer}, {Wik}, {Hornschemeier},
  {Ptak}, {Antoniou}, {Argo}, {Bechtol}, {Boggs}, {Christensen}, {Craig},
  {Hailey}, {Harrison}, {Krivonos}, {Leyder}, {Maccarone}, {Stern}, {Venters},
  {Zezas}, \& {Zhang}}]{Lehmer+13}
{Lehmer} B.~D. {et~al.}, 2013, \apj, 771, 134

\bibitem[{{Lenc} \& {Tingay}(2006)}]{Lenc+06}
{Lenc} E., {Tingay} S.~J., 2006, \aj, 132, 1333

\bibitem[{{Leroy} {et~al}\mbox{.}(2015){Leroy}, {Bolatto}, {Ostriker},
  {Rosolowsky}, {Walter}, {Warren}, {Donovan Meyer}, {Hodge}, {Meier}, {Ott},
  {Sandstrom}, {Schruba}, {Veilleux}, \& {Zwaan}}]{Leroy+15}
{Leroy} A.~K. {et~al.}, 2015, \apj, 801, 25

\bibitem[{{Leurini} {et~al}\mbox{.}(2016){Leurini}, {Menten}, \&
  {Walmsley}}]{Leurini+16}
{Leurini} S., {Menten} K.~M., {Walmsley} C.~M., 2016, \aap, 592, A31

\bibitem[{{Levshakov} {et~al}\mbox{.}(2011){Levshakov}, {Kozlov}, \&
  {Reimers}}]{Levshakov+11}
{Levshakov} S.~A., {Kozlov} M.~G., {Reimers} D., 2011, \apj, 738, 26

\bibitem[{{Mart{\'{\i}}n} {et~al}\mbox{.}(2006{\natexlab{a}}){Mart{\'{\i}}n},
  {Mart{\'{\i}}n-Pintado}, \& {Mauersberger}}]{Martin+06b}
{Mart{\'{\i}}n} S., {Mart{\'{\i}}n-Pintado} J., {Mauersberger} R.,
  2006{\natexlab{a}}, \aap, 450, L13

\bibitem[{{Mart{\'{\i}}n} {et~al}\mbox{.}(2005){Mart{\'{\i}}n},
  {Mart{\'{\i}}n-Pintado}, {Mauersberger}, {Henkel}, \&
  {Garc{\'{\i}}a-Burillo}}]{Martin+05}
{Mart{\'{\i}}n} S., {Mart{\'{\i}}n-Pintado} J., {Mauersberger} R., {Henkel} C.,
  {Garc{\'{\i}}a-Burillo} S., 2005, \apj, 620, 210

\bibitem[{{Mart{\'{\i}}n} {et~al}\mbox{.}(2006{\natexlab{b}}){Mart{\'{\i}}n},
  {Mauersberger}, {Mart{\'{\i}}n-Pintado}, {Henkel}, \&
  {Garc{\'{\i}}a-Burillo}}]{Martin+06}
{Mart{\'{\i}}n} S., {Mauersberger} R., {Mart{\'{\i}}n-Pintado} J., {Henkel} C.,
  {Garc{\'{\i}}a-Burillo} S., 2006{\natexlab{b}}, \apjs, 164, 450

\bibitem[{{Matsubayashi} {et~al}\mbox{.}(2009){Matsubayashi}, {Sugai},
  {Hattori}, {Kawai}, {Ozaki}, {Kosugi}, {Ishigaki}, \&
  {Shimono}}]{Matsubayashi+09}
{Matsubayashi} K., {Sugai} H., {Hattori} T., {Kawai} A., {Ozaki} S., {Kosugi}
  G., {Ishigaki} T., {Shimono} A., 2009, \apj, 701, 1636

\bibitem[{{McEwen} {et~al}\mbox{.}(2014){McEwen}, {Pihlstr{\"o}m}, \&
  {Sjouwerman}}]{McEwen+14}
{McEwen} B.~C., {Pihlstr{\"o}m} Y.~M., {Sjouwerman} L.~O., 2014, \apj, 793, 133

\bibitem[{{McEwen} {et~al}\mbox{.}(2016){McEwen}, {Sjouwerman}, \&
  {Pihlstr{\"o}m}}]{McEwen+16}
{McEwen} B.~C., {Sjouwerman} L.~O., {Pihlstr{\"o}m} Y.~M., 2016, \apj, 832, 129

\bibitem[{{Meier} \& {Turner}(2005)}]{Meier+05}
{Meier} D.~S., {Turner} J.~L., 2005, \apj, 618, 259

\bibitem[{{Meier} \& {Turner}(2012)}]{Meier+12}
{Meier} D.~S., {Turner} J.~L., 2012, \apj, 755, 104

\bibitem[{{Meier} {et~al}\mbox{.}(2015){Meier}, {Walter}, {Bolatto}, {Leroy},
  {Ott}, {Rosolowsky}, {Veilleux}, {Warren}, {Wei{\ss}}, {Zwaan}, \&
  {Zschaechner}}]{Meier+15}
{Meier} D.~S. {et~al.}, 2015, \apj, 801, 63

\bibitem[{{Menten} {et~al}\mbox{.}(1992){Menten}, {Reid}, {Pratap}, {Moran}, \&
  {Wilson}}]{Menten+92}
{Menten} K.~M., {Reid} M.~J., {Pratap} P., {Moran} J.~M., {Wilson} T.~L., 1992,
  \apjl, 401, L39

\bibitem[{{M{\"u}ller} {et~al}\mbox{.}(2004){M{\"u}ller}, {Menten}, \&
  {M{\"a}der}}]{Muller+04}
{M{\"u}ller} H.~S.~P., {Menten} K.~M., {M{\"a}der} H., 2004, \aap, 428, 1019

\bibitem[{{M{\"u}ller-S{\'a}nchez}
  {et~al}\mbox{.}(2010){M{\"u}ller-S{\'a}nchez}, {Gonz{\'a}lez-Mart{\'{\i}}n},
  {Fern{\'a}ndez-Ontiveros}, {Acosta-Pulido}, \& {Prieto}}]{Muller-Sanchez+10}
{M{\"u}ller-S{\'a}nchez} F., {Gonz{\'a}lez-Mart{\'{\i}}n} O.,
  {Fern{\'a}ndez-Ontiveros} J.~A., {Acosta-Pulido} J.~A., {Prieto} M.~A., 2010,
  \apj, 716, 1166

\bibitem[{{Ott} {et~al}\mbox{.}(2005){Ott}, {Weiss}, {Henkel}, \&
  {Walter}}]{Ott+05}
{Ott} J., {Weiss} A., {Henkel} C., {Walter} F., 2005, \apj, 629, 767

\bibitem[{{Phillips} {et~al}\mbox{.}(1998{\natexlab{a}}){Phillips}, {Norris},
  {Ellingsen}, \& {McCulloch}}]{Phillips+98b}
{Phillips} C.~J., {Norris} R.~P., {Ellingsen} S.~P., {McCulloch} P.~M.,
  1998{\natexlab{a}}, \mnras, 300, 1131

\bibitem[{{Phillips} {et~al}\mbox{.}(1998{\natexlab{b}}){Phillips}, {Norris},
  {Ellingsen}, \& {Rayner}}]{Phillips+98a}
{Phillips} C.~J., {Norris} R.~P., {Ellingsen} S.~P., {Rayner} D.~P.,
  1998{\natexlab{b}}, \mnras, 294, 265

\bibitem[{{Pihlstr{\"o}m} {et~al}\mbox{.}(2014){Pihlstr{\"o}m}, {Sjouwerman},
  {Frail}, {Claussen}, {Mesler}, \& {McEwen}}]{Pihlstrom+14}
{Pihlstr{\"o}m} Y.~M., {Sjouwerman} L.~O., {Frail} D.~A., {Claussen} M.~J.,
  {Mesler} R.~A., {McEwen} B.~C., 2014, \aj, 147, 73

\bibitem[{{Plambeck} \& {Menten}(1990)}]{Plambeck+90}
{Plambeck} R.~L., {Menten} K.~M., 1990, \apj, 364, 555

\bibitem[{{Radovich} {et~al}\mbox{.}(2001){Radovich}, {Kahanp{\"a}{\"a}}, \&
  {Lemke}}]{Radovich+01}
{Radovich} M., {Kahanp{\"a}{\"a}} J., {Lemke} D., 2001, \aap, 377, 73

\bibitem[{{Rosenberg} {et~al}\mbox{.}(2014){Rosenberg}, {Kazandjian}, {van der
  Werf}, {Israel}, {Meijerink}, {Wei{\ss}}, {Requena-Torres}, \&
  {G{\"u}sten}}]{Rosenberg+14}
{Rosenberg} M.~J.~F., {Kazandjian} M.~V., {van der Werf} P.~P., {Israel} F.~P.,
  {Meijerink} R., {Wei{\ss}} A., {Requena-Torres} M.~A., {G{\"u}sten} R., 2014,
  \aap, 564, A126

\bibitem[{{Sakamoto} {et~al}\mbox{.}(2006){Sakamoto}, {Ho}, {Iono}, {Keto},
  {Mao}, {Matsushita}, {Peck}, {Wiedner}, {Wilner}, \& {Zhao}}]{Sakamoto+06}
{Sakamoto} K. {et~al.}, 2006, \apj, 636, 685

\bibitem[{{Sakamoto} {et~al}\mbox{.}(2011){Sakamoto}, {Mao}, {Matsushita},
  {Peck}, {Sawada}, \& {Wiedner}}]{Sakamoto+11}
{Sakamoto} K., {Mao} R.-Q., {Matsushita} S., {Peck} A.~B., {Sawada} T.,
  {Wiedner} M.~C., 2011, \apj, 735, 19

\bibitem[{{Sjouwerman} {et~al}\mbox{.}(2010){Sjouwerman}, {Murray},
  {Pihlstr{\"o}m}, {Fish}, \& {Araya}}]{Sjouwerman+10}
{Sjouwerman} L.~O., {Murray} C.~E., {Pihlstr{\"o}m} Y.~M., {Fish} V.~L.,
  {Araya} E.~D., 2010, \apjl, 724, L158

\bibitem[{{Turner}(1985)}]{Turner+85}
{Turner} B.~E., 1985, \apj, 299, 312

\bibitem[{{Ulvestad} \& {Antonucci}(1997)}]{Ulvestad+97}
{Ulvestad} J.~S., {Antonucci} R.~R.~J., 1997, \apj, 488, 621

\bibitem[{{Voronkov} {et~al}\mbox{.}(2006){Voronkov}, {Brooks}, {Sobolev},
  {Ellingsen}, {Ostrovskii}, \& {Caswell}}]{Voronkov+06}
{Voronkov} M.~A., {Brooks} K.~J., {Sobolev} A.~M., {Ellingsen} S.~P.,
  {Ostrovskii} A.~B., {Caswell} J.~L., 2006, \mnras, 373, 411

\bibitem[{{Voronkov} {et~al}\mbox{.}(2014){Voronkov}, {Caswell}, {Ellingsen},
  {Green}, \& {Breen}}]{Voronkov+14}
{Voronkov} M.~A., {Caswell} J.~L., {Ellingsen} S.~P., {Green} J.~A., {Breen}
  S.~L., 2014, \mnras, 439, 2584

\bibitem[{{Wang} {et~al}\mbox{.}(2014){Wang}, {Zhang}, {Gao}, {Zhang}, {Li},
  {Fang}, \& {Shi}}]{Wang+14}
{Wang} J., {Zhang} J., {Gao} Y., {Zhang} Z.-Y., {Li} D., {Fang} M., {Shi} Y.,
  2014, Nature Communications, 5, 5449

\bibitem[{{Wei{\ss}} {et~al}\mbox{.}(2008){Wei{\ss}}, {Kov{\'a}cs},
  {G{\"u}sten}, {Menten}, {Schuller}, {Siringo}, \& {Kreysa}}]{Weiss+08}
{Wei{\ss}} A., {Kov{\'a}cs} A., {G{\"u}sten} R., {Menten} K.~M., {Schuller} F.,
  {Siringo} G., {Kreysa} E., 2008, \aap, 490, 77

\bibitem[{{Wilson} {et~al}\mbox{.}(2011){Wilson}, {Ferris}, {Axtens}, {Brown},
  {Davis}, {Hampson}, {Leach}, {Roberts}, {Saunders}, {Koribalski}, {Caswell},
  {Lenc}, {Stevens}, {Voronkov}, {Wieringa}, {Brooks}, {Edwards}, {Ekers},
  {Emonts}, {Hindson}, {Johnston}, {Maddison}, {Mahony}, {Malu}, {Massardi},
  {Mao}, {McConnell}, {Norris}, {Schnitzeler}, {Subrahmanyan}, {Urquhart},
  {Thompson}, \& {Wark}}]{Wilson+11}
{Wilson} W.~E. {et~al.}, 2011, \mnras, 416, 832

\bibitem[{{Yusef-Zadeh} {et~al}\mbox{.}(2013){Yusef-Zadeh}, {Cotton}, {Viti},
  {Wardle}, \& {Royster}}]{Yusef-Zadeh+13}
{Yusef-Zadeh} F., {Cotton} W., {Viti} S., {Wardle} M., {Royster} M., 2013,
  \apjl, 764, L19

\end{thebibliography}

%\appendix

\end{document}